\documentclass[aps,prc,twocolumn,showpacs,floatfix]{revtex4}
\usepackage[dvips]{graphicx}
\usepackage{bm}

\def\bea{\begin{eqnarray}}
\def\eea{\end{eqnarray}}
\def\be{\begin{equation}}
\def\ee{\end{equation}}
\def\rra{\right\rangle}
\def\lla{\left\langle}

\begin{document}
\title{Non-locality in the nucleon-nucleon interaction and 
nuclear matter saturation}
\author{M. Baldo and C. Maieron}
\affiliation{ 
INFN, Sezione di Catania, Via Santa Sofia 64, 95123 Catania, Italy}

\date{\today} 

\begin{abstract}
We study the possible relationship between the saturation properties of
nuclear matter
and the inclusion of non-locality in the nucleon-nucleon 
interaction. To this purpose we compute the saturation curve of
nuclear matter within the  Bethe-Brueckner-Goldstone theory
using a recently proposed realistic non-local potential, and compare
it with the corresponding curves obtained with a purely local realistic
interaction (Argonne v$_{18}$) and the most recent version of the
one-boson exchange potential
(CD Bonn). We find that the inclusion of non-locality in the two-nucleon
bare interaction strongly affects saturation, but it is unable to provide
a consistent description of few-body nuclear systems and nuclear matter.
 
\end{abstract}
\pacs{
      21.65.+f,  
      24.10.Cn,  
      21.30.-x   
 }
\maketitle

\section{Introduction}
The effective nucleon-nucleon (NN) interaction is expected, on physical
grounds, to be intrinsically non-local. The quark structure of nucleons 
implies that the potential between two interacting nucleons
cannot be expressed only in terms of the center of mass degree of freedom,
at least for distances smaller than twice the nucleon radius. 
Even a description of the NN interaction in terms of one-boson
exchange processes introduces some degree of non-locality \cite{obe}.
For practical purposes, however, it is convenient to have a local
(energy independent) NN interaction, to be used in numerical applications.
The modeling of the NN interaction by a local potential, with possible
guidance from meson exchange processes, has been extremely successful
in reproducing the NN experimental phase shifts and deuteron
properties \cite{phase1,phase2}. This approach has developed for many years
and has reached a high degree of sophistication. \par
It has been established by now that these local NN potentials,
which are essentially phase equivalent, give slightly different results 
for the three- and four-nucleon systems. Furthermore, in general they 
underestimate the binding
energy and do not give the correct values for some polarization
observables. This drawback has been usually overcome by introducing
three-body forces, which can be again justified by the non-elementary
nature of the nucleon. Actually three-body forces can be generated
also by meson-exchange processes of higher order, where the nucleon
is excited in the intermediate states of the interaction processes or
meson-meson couplings are introduced \cite{threebody}. Both non-locality
and three-body forces can be considered to have a common origin, the 
inner structure of the nucleon. However, one cannot fully reduce  
one to another, since non-locality has definite effects already
at the two-body level, e.g. a non-local potential cannot be equivalent, 
in general, to an energy independent local potential. Furthermore,
it is not clear to what extent the off-shell behavior of a non-local
potential, i.e. the off shell $T$-matrix, can be simulated by the
presence of three-body forces.\par
Another question where non-locality could play a role is the mechanism
of saturation in nuclear matter. In fact, it turns out that in order
to get saturation with a local (energy independent) interaction, 
it is essential to include a strong repulsive core, as also suggested
by the behavior of the NN phase shifts.
Again, three-body forces are necessary to tune the saturation point 
close to the empirical one.
It is surely conceivable that a certain degree of non-locality could
simulate the strong repulsion at short distance, which is one of the main
features of the NN interaction, and maybe reduce the need of three-body
forces. \par
It appears, therefore, of great interest to investigate to what extent
non-locality alone could enable to get saturation, possibly close to the
empirical one, and at the same time to describe  three- and four-body
systems correctly.\par
Recently, in a series of papers 
by Doleschall and collaborators
\cite{Doleschall:2000rp, Doleschall:2003ha,Doleschall:2004bb},
a new realistic non-local NN potential
was proposed, which is able to describe accurately both the NN phase
shifts and the properties of three-nucleon
systems ($^3He$ and $^3H$), i.e. their binding energies and radii.
Also the alpha particle appears to be reasonably well described 
\cite{Lazauskas:2004hq}.
The need of three-body forces in this case seems to be absent, 
or completely simulated by the non-locality. \par
This NN interaction introduces non-locality at short distance 
($r \leq 3$ fm$^{-1}$) for the s-wave channels, and possibly also for 
the p-waves ones \cite{Doleschall:2004bb}. 
The rest of the potential is taken as the  Argonne v$_{18}$
\cite{phase1},
while the non-local part 
is purely phenomenological.\par
In this paper we address the issue of the saturation properties
of non-local potentials.
To this purpose
we present nuclear matter calculations within the Bethe-Brueckner-Goldstone
method extended up to three hole-line contributions, which is known
to converge well and to give an accurate nuclear matter saturation curve.\par
As the main representative of non-local NN interaction
we take the potential of ref. \cite{Doleschall:2003ha}. 
We compare the corresponding nuclear matter Equation of state (EOS)
with those obtained starting from the charge-dependent
(CD) Bonn \cite{Machleidt:2000ge} and the v$_{18}$ potentials.
The former can be considered
the most recent and accurate interaction based
on the one-boson exchange model
and, therefore, it contains some degree of non-locality.
The latter is a convenient reference local potential for the comparison
with Doleschall's, which, as we stated above, 
coincides with the v$_{18}$ in the $L\ge 1$ channels.
Additionally, other modern local potentials, 
like Argonne v$_{14}$ \cite{phase2}
or Paris
potential \cite{paris}, give similar results, 
close enough to 
v$_{18}$ to be considered equivalent for the present purpose 
\cite{book}.\par
The study of phase-shift equivalent NN interactions and their predictions
of nuclear matter saturation properties has a long history. For example
in refs.~\cite{Coester,Sauer} unitarily equivalent interactions which 
differ in the short range part of the $^1S_0$ channel were shown to
provide rather different values of the saturation energy and density.
More recently, the predictions of modern NN potentials were studied
in  refs.~\cite{Engvik:1997er,mach98,mach96}. Here the role
of non-locality in the binding energy of nuclear matter
was also partially discussed, using
the CD Bonn and the Nijmegen-I \cite{nijmI} potentials, which were
shown to provide  a stronger binding than local potentials such as
the v$_{18}$.
However, both these potentials underestimate the binding energy 
of three- and four-body systems \cite{cd_few}, at variance with
Doleschall's interaction \cite{Doleschall:2003ha}. 
Therefore when using the latter
we can, in principle, expect different
results also for nuclear matter calculations. \par
It is well known that the saturation points obtained within
the Brueckner approximation for the gap choice and different NN interactions
tend to lie along a band \cite{book},
the celebrated ``Coester band", in the energy-density plane.
The position inside the band for a given NN interaction
turns out to be correlated with the d-state
probability in the deuteron.
The high-energy, high-density
part of the band is usually associated with low values of this probability.
However, as shown in ref. \cite{BDay}, this trend does not hold any more
if three hole-line contributions are included, as we do in the present
paper. In this case the results are quite insensitive to the d-wave
probability value and actually close to each other, at least for local
potentials or quasi-local potentials like the Paris one. The same conclusion
holds true if the continuous choice for the single particle potential is used
in the Brueckner approximation \cite{book}.

The paper is organized in the following way. In Section \ref{s:pot}
we shortly study the free space properties of the Doleschall, CD Bonn and 
v$_{18}$
potentials. In Section \ref{s:eos}, after introducing the BBG formalism,
we study the EOS of nuclear matter obtained with Doleschall's potential,
addressing, in particular, the issue of the convergence of the hole-line 
expansion in this case. We then compare this EOS with those obtained
starting from the two other potentials. Finally in Section \ref{s:concl}
we draw our conclusions.
 
\section{Local vs. non-local NN potentials. }
\label{s:pot}
\begin{figure} 
\includegraphics[width=0.95\columnwidth]{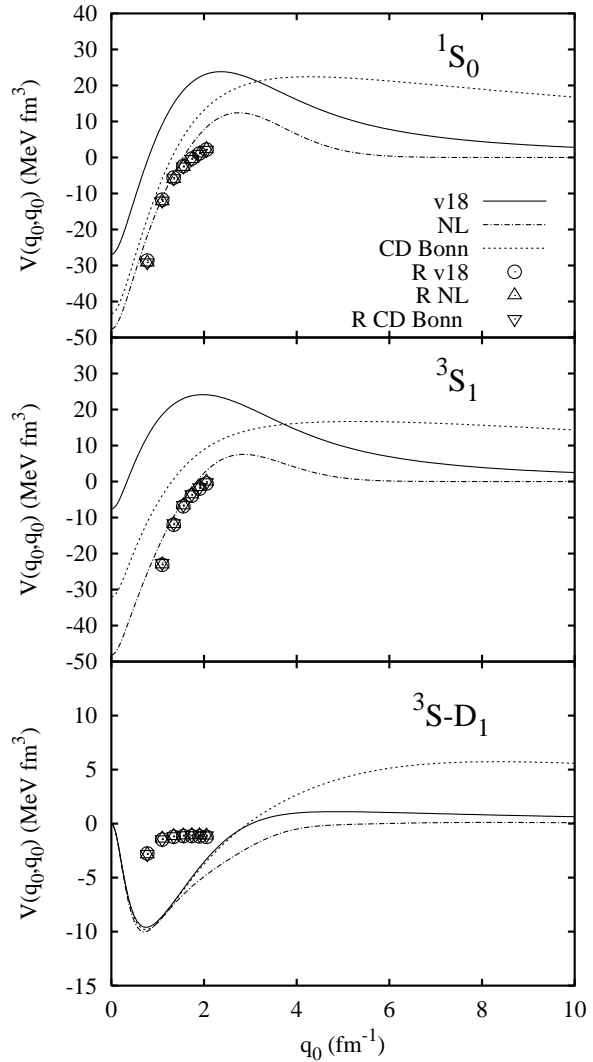}
\caption{
Diagonal matrix elements of the v$_{18}$ 
(solid line), CD Bonn (dotted) potentials
and of Doleshall's IS potential of ref.~\cite{Doleschall:2003ha} 
(dot-dashed, labeled as NL), 
for the np interaction
in the $^1S_0$ (upper panel),  $^3S_1$ (middle) and in the
coupling $^3S_1$-$^3D_1$ (lower panel) channels.
The on-shell $R(q_0,q_0;E)$ $R$-matrix elements are also shown (symbols), 
for laboratory energies ranging  from $50$ to $350$
MeV, in steps of 50 MeV.}  
\label{fig:diag}
\end{figure} 
Before studying infinite nuclear matter
and in order to better understand the role played by the different
NN interactions in the nuclear matter calculations of next section, 
it is useful
to briefly compare the properties and characteristics
of the potentials we are going to employ, on the basis also of some
established results obtained in the literature 
\cite{phase1,Machleidt:2000ge,Engvik:1997er}
%
\begin{figure*} 
\includegraphics[width=0.8\textwidth]{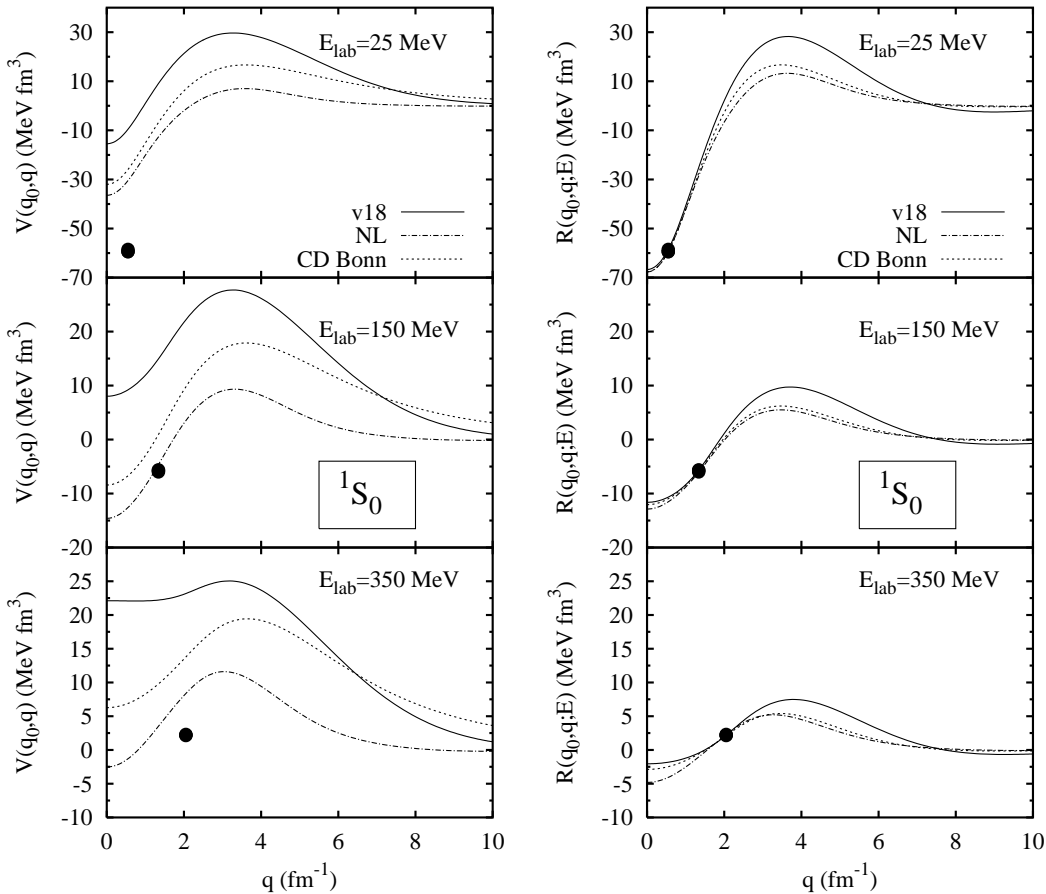}
\caption{
Left panels: off-diagonal $^1S_0$ $np$
matrix elements of the v$_{18}$, CD Bonn and 
NL potentials,
for three fixed momenta $q_0$ corresponding to laboratory energies
$E_{lab}= 2 E$ = $25$ (upper panel), $150$ (middle) and $350$ (lower)
MeV.
Right panels: corresponding half off-shell $R$-matrix elements.
The solid dot marks the on-shell values of the $R$-matrix.}
\label{fig:half1s0}
\end{figure*} 
%
\begin{figure*} 
\includegraphics[width=0.8\textwidth]{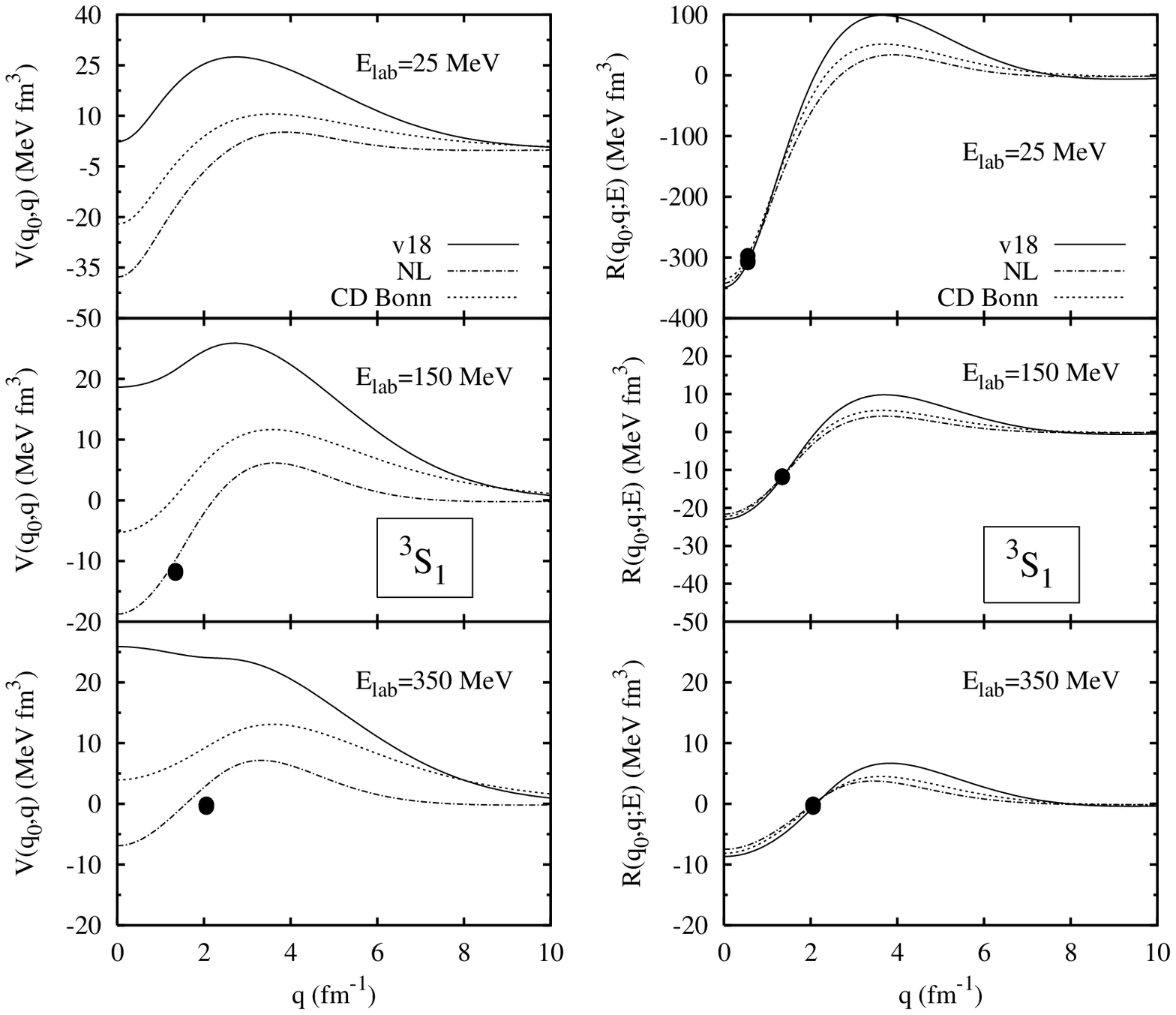}
\caption{
As fig.~\ref{fig:half1s0} but for the $^3S_1$ matrix elements.}
\label{fig:half3s1}
\end{figure*} 
%
\par
The general matrix elements in the momentum representation of 
a NN interaction $V^c(r,r')$ in a given two-body channel $c$
can be written
\be
 V^c_{ll'}(k,k') \, =\, 
{2\over\pi} \int r^2 dr r'^2 dr' j_l(kr) V^c_{ll'}(r,r')
                                           j_l'(k'r') 
\label{eq:v}
\ee
where $l,l'$ are the initial and final orbital angular momenta in the
channel, and $j_l , j_l'$ the corresponding spherical Bessel functions.
For a local potential $V^c (r,r') = v^c \delta(r - r')/rr'$.
In the momentum representation the matrix elements $V^c_{ll'}(k,k')$
of eq. (\ref{eq:v}) are in any case a function of two variables, the
initial and final relative momenta  $k$ and $k'$. 
The non-locality is incorporated in a non trivial way in
the analytical properties of the matrix elements of the potentials,
as well as of the corresponding
$T$-Matrix, or of the real $R$-Matrix, which is used here
for later convenience.
The latter describes the free space NN scattering and
is obtained by solving the Lippmann-Schwinger equation, which
for two nucleons interacting with energy $E$ in the center of mass 
frame can be written as:
\bea
&& R^c_{ll'}(k,k';E) = V^c_{ll'}(k,k') + 
\label{eq:lippmann}
\\
&& \,\,\, + \displaystyle{
\sum_{l^{\prime\prime}}
{\mathcal{P}} \int_0^\infty q^2 dq V^c_{l l^{\prime\prime}}(k,q)
\frac{M}{M E -q^2} R^c_{l^{\prime\prime}l'}(q,k';E)}\,,
\nonumber
\eea
where the symbol ${\mathcal{P}}$ denotes the principal value integral
and $M$ is the free nucleon mass.

For realistic potentials the on-shell $R$-matrix, 
$R(q_0,q_0;E)$ with $q_0=\displaystyle{\sqrt{M E}}$, is fixed, to a large
extent, by the fitting of the experimental NN phase shifts. 

The potentials we consider in the present paper were fit 
to reproduce the Nijmegen phase shifts \cite{Stoks:1993tb} for 
 energies up to $350$ MeV in the laboratory frame, and therefore
give rise to practically identical on-shell values of the $R$-matrix
in this energy range. However, due to their totally different analytical
structure, 
a quite
different behavior of the bare NN matrix elements of eq. (\ref{eq:v})
can be expected, especially when non-local potentials are compared with
local ones.

This is illustrated in fig.~\ref{fig:diag}, 
where the bare NN diagonal 
matrix elements and the corresponding
on-shell $R$-matrix for the local v$_{18}$ and the non-local 
CD Bonn are compared with those obtained from the 
non-local potential indicated as IS  in ref.~\cite{Doleschall:2003ha}
and hereafter referred to as NL.
Here the dominant S-wave channels are considered, 
for which the main non-locality 
has been introduced in ref. \cite{Doleschall:2003ha}. 
One can see that, despite the expected agreement of the various interactions 
on the $R$-matrix, the behavior of the bare NN matrix elements is 
quite different.
At small momenta we see that in the $^1S_0$ channel
the two non-local potentials are rather close to each other,
while they differ significantly from the v$_{18}$ This is still
true in the $^3S_1$ channel, where, however, the difference between
NL and CD Bonn is increased. 
It can be also seen that
at increasing non-locality, i.e. going from the CD Bonn to the NL
potential, the deviation form the local potential v$_{18}$ is increasing.
This is in agreement with refs.~\cite{Engvik:1997er,mach98}, 
where it was shown
that including non-locality makes the attraction provided by 
the integral term of eq.~(\ref{eq:lippmann}) weaker,
thus requiring a less repulsive diagonal matrix element
of $V$, in order to obtain the same on-shell $R$-matrix.
In particular it is interesting to note that 
the diagonal NL potential is always
rather close to the corresponding on-shell $R$-matrix elements,
especially for the $^3S_1$ channel.
In the coupling channel $^3S_1$-$^3D_1$ we see that the three
potentials are very close to each other at small momenta, with 
NL deviating from the others at intermediate $q$. 

Considering the large momentum behavior, we observe that in all cases
the CD Bonn potentials exhibits a much larger tail.
\par
Since the diagonal matrix elements of the bare potentials
are already so different from each other, is not surprising then
that the off-shell behavior of both the bare $V$ and 
the $R$-matrix is also different,
 as illustrated in 
figs.~\ref{fig:half1s0} and \ref{fig:half3s1}, where the half off-shell
matrix elements of the bare potentials and of the $R$-matrix are shown
for three choices of the incident energy. The behavior of the bare 
potentials (left panels in the figures) shows essentially the same features
we have already observed for the diagonal matrix elements.
The curves representing the $R$-matrix (right panels) are closer 
to each other, but still sensible differences are observed, especially
when the results corresponding to the v$_{18}$ are compared with
the two other non-local potentials.
\begin{figure*} 
\includegraphics[width=0.8\textwidth]{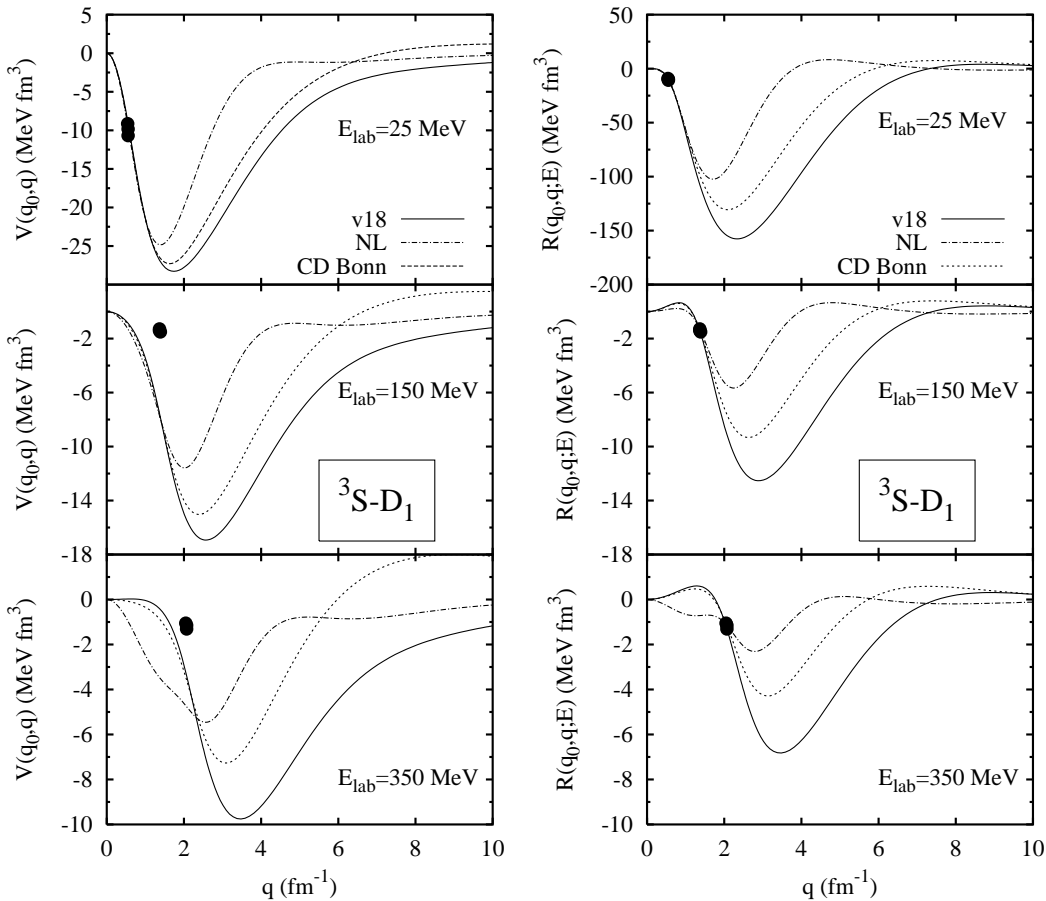}
\caption{
As fig.~\ref{fig:half1s0} but for the $^3S_1-^3D_1$ matrix elements.}
\label{fig:half3sd1}
\end{figure*} 
\par
Finally in fig.~\ref{fig:half3sd1} we also show the half off-shell
$V$ and $R$  $^3S$-$D_1$ matrix elements. Here only the tensor
part of the NN interaction can contribute, which was shown
in previous works to be strongly affected by the inclusion of
non-locality ~\cite{Engvik:1997er,mach98,mach96}, 
and which is known to play an important role
in nuclear structure calculations.
We see that indeed, the inclusion of non-locality reduces the
strength of the interaction and, for the NL case, also the range
of momentum values where the matrix elements are significantly different
from zero.
\section{EOS of nuclear matter}
\label{s:eos}
\subsection{The BBG expansion}
\label{s:bbg}
The microscopic calculations of nuclear matter Equation of State, i.e.
the energy per particle as a function of density, on the basis of realistic
interactions has a long history. Theoretical and numerical methods, 
accurate enough for our considerations, are nowadays available. We will
follow the Bethe-Brueckner-Goldstone scheme, which has been proved 
\cite{book} to be reliable in a wide range of density. 
In this expansion the
original bare NN interaction $v$ is systematically replaced by the
$G$-matrix, which satisfies the integral equation
\be
 G[\rho;\omega] = v + \sum_{k_a, k_b} v 
 { \left|k_a k_b\rra Q \lla k_a k_b\right|
 \over \omega - e(k_a) - e(k_b) } G[\rho;\omega] \:, 
\label{eq:bg}
\ee
where $e(k)$ is the single particle energy and $Q$ the so-called Pauli
operator which projects both intermediate single particle states above
the Fermi surface. The diagrams in terms of the $G$-matrix are then 
ordered according to the numbers of hole-lines they contain. The 
introduction of the self-consistent single particle potential $U(k)$
is an essential ingredient of the method 
\be
 U(k) = \sum_{k'<k_F} \lla k k' | G (e(k)+e(k'))| k k'\rra_A 
\label{eq:u}
\ee
Since $e(k) = \hbar^2 k^2/2M + U(k)$, eqs. (\ref{eq:bg},\ref{eq:u}) 
include a self-consistent procedure which determines $U(k)$. 
In particular we use the continuous choice for the potential $U(k)$, i.e.
the definition of eq. (\ref{eq:u}) is extended to all momenta $k$, below and 
above the Fermi momentum $k_F$. More details can be found in review 
papers 
or in textbooks \cite{book}. 
We extend our calculations 
up to the three-hole line level of approximation, which has been
shown to be accurate in 
refs.~\cite{Baldo2002,song,song1} 
for a variety of NN potentials,
including the Argonne v$_{14}$ and v$_{18}$ and the Paris interactions.

\begin{figure}[t] 
\includegraphics[width=0.95\columnwidth]{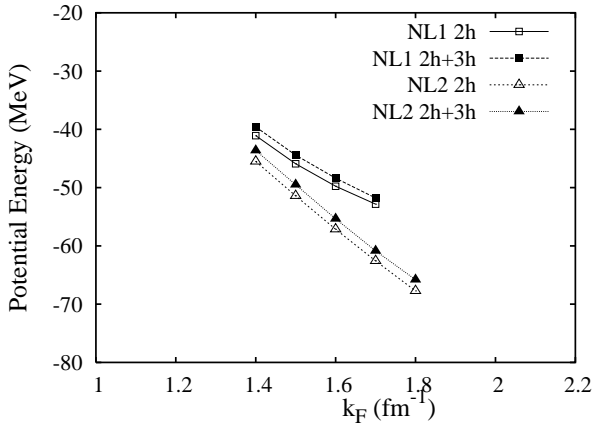}
\caption{
Potential energy per particle for symmetric nuclear matter calculated
with the non-local potentials NL1 (squares), which includes the non-locality
in the $^1S_0$ channel only, and NL2 (triangles), which has the non-locality
both in the $^1S_0$ and in the $^3S-D_1$ channels.
Empty symbols correspond to the two hole-line calculations, while full symbols
represent the full two + three hole-line results.}
\label{fig:bind}
\end{figure}

However, since Doleschall's NL potential is used here for the first
time in BBG calculations of nuclear matter, before making comparisons
of different EOS, we briefly study the accuracy of the BBG expansion
in this case. This is done 
in fig.~\ref{fig:bind}, where the contributions to the binding potential
energy
of the two hole-line (Brueckner) and three hole-line diagrams are reported.
To keep the calculations reliable, the latter has to be substantially 
smaller than the former. \par
Two different choices for the NL potential
are considered:
the curves labeled as NL1 have been obtained using Doleschall's  IS
potential of ref.~\cite{Doleschall:2003ha}
only in the $^1S_0$ channel, while for the curves labeled
as NL2 the non-local potential has been included also in the 
$^3S_1$-$^3D_1$ channel. In the remaining channels the v$_{18}$ 
interaction has been used.

The three hole-line contributions turn out to be slightly larger than in
the case of local potentials (see \cite{Baldo2002,song,song1}).
However these corrections always remain within $2$-$3.5$\% for 
the NL1 case and within $2.5$-$4$\% for NL2, thus indicating that
our calculations up to three hole-lines can be considered accurate.
It is interesting to note that with the non-local potentials 
the corrections to the BHF curves are positive, 
contrary to the case of the  v$_{18}$, where, in the region
$1.4 < k_F < 2$ $fm~^{-1}$ the three hole-line contributions tend
to make the EOS more attractive.

\subsection{Nuclear matter saturation properties}
\label{s:sat}
Having established the validity of our calculations, we can now turn
our attention to the study of the saturation properties of nuclear
matter. To better understand the role that the different potentials
we studied in section \ref{s:pot} play in the nuclear matter EOS
it is useful to rewrite the Bethe-Goldstone equation~(\ref{eq:bg})
in the channel $c$,
\bea
&& \langle k l | G^c(\rho,\omega) | k' l'\rangle = V^c_{l l'}(k,k')
\label{eq:bg_c}
\\
&& \,\,\, + \displaystyle{
\sum_{l^{\prime\prime}}
\int_0^\infty q^2 dq V^c_{l l^{\prime\prime}}(k,q)
\frac{\overline{Q}(q)}{\omega -\overline{E}(q)} 
\langle q l^{\prime\prime}  |G^c(\rho,\omega) | k' l'\rangle}\,,
\nonumber
\eea
where, following standard procedures \cite{book}, we have introduced
the angle averaged
expressions of the Pauli operator, $\overline{Q}$, and of the energy,
$\overline{E}$, in the denominator of the integral term. 
Eq.~(\ref{eq:bg_c}) closely resembles the Lippmann-Schwinger
equation ~(\ref{eq:lippmann}), the difference being due to
the presence, in the former, 
of the Pauli blocking operator and of a different energy denominator.
Both these terms produce a quenching of the integral term contribution
in the Bethe-Goldstone equation with respect to the Lippmann-Schwinger
case.
Therefore the behavior of the G-matrix is expected to be rather similar
to that of the R-matrix, with this quenching effect being weaker for
the non-local potentials, for which the integral term is already small.
\par
\begin{figure}[t] 
\includegraphics[width=0.95\columnwidth]{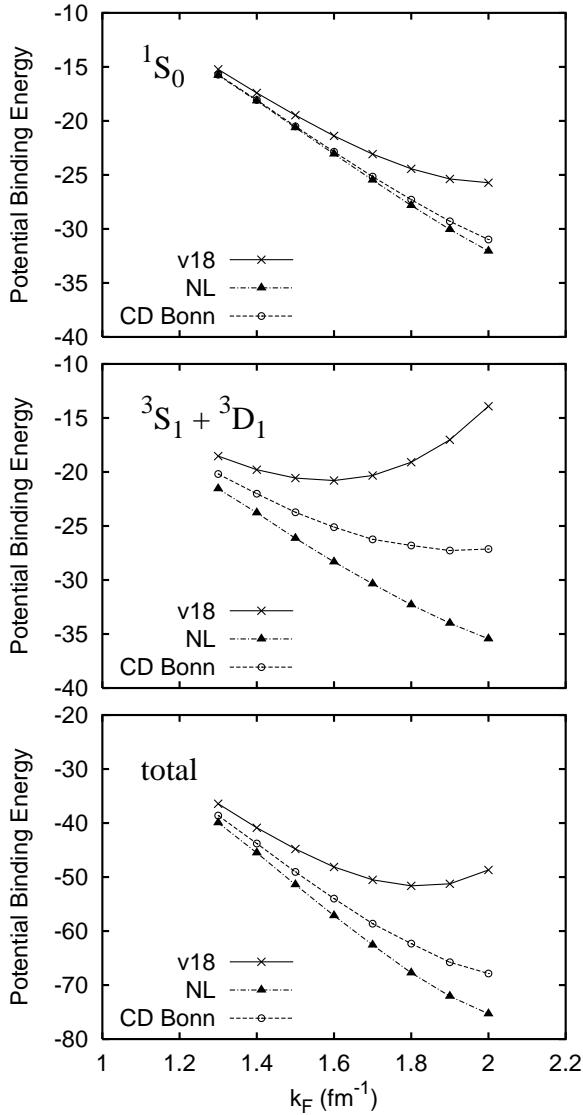}
\caption{
Contributions of the different channels to the potential binding 
energy per particle
(in MeV) vs. the Fermi momentum $k_F$. The upper panel corresponds to
the $^1S_0$ channel, the middle panel to the $^3S_1 + ^3D_1$ channels,
while the lower panel shows the total potential binding energy.
All curves are calculated at the 2 hole-line level.}
\label{fig:bind_can}
\end{figure} 
Since the non-locality in the NL potential is introduced in the
$^1S_0$ and in the coupled $^3S_1$-$^3D_1$ channels, we begin by studying
the contributions of these channels to the  potential binding energy
per particle. This is done, at the Brueckner level, in 
fig.~\ref{fig:bind_can}, where the results corresponding to the non-local
CD Bonn (empty dots/dashed line)  
and NL potentials (triangles/dot-dashed) are compared with
the v$_{18}$ (crosses/solid). 
In agreement with the discussion in section~\ref{s:pot}, where we 
compared bare potentials and R-matrix elements,
we observe that in the
$^1S_0$ channel the CD Bonn and NL potential contributions
are very close to each other and that they provide additional
binding, with respect to the v$_{18}$ case. 
When we consider the $^3S_1$ and $^3D_1$ channels,
again we see that the non-local potentials provide more binding, but
in this case the NL curve is significantly lower than the CD Bonn one.
Since the contributions from all other channels  
are very similar for the three potentials (with differences of the order 
of 1 MeV at the largest densities considered here), the differences
observed in the  $^1S_0$ and  $^3S_1$ and $^3D_1$ channels determine
the behavior of the total binding potential energy, shown in the bottom panel 
of fig.~\ref{fig:bind_can}.
\par
\begin{figure} 
\includegraphics[width=0.92\columnwidth]{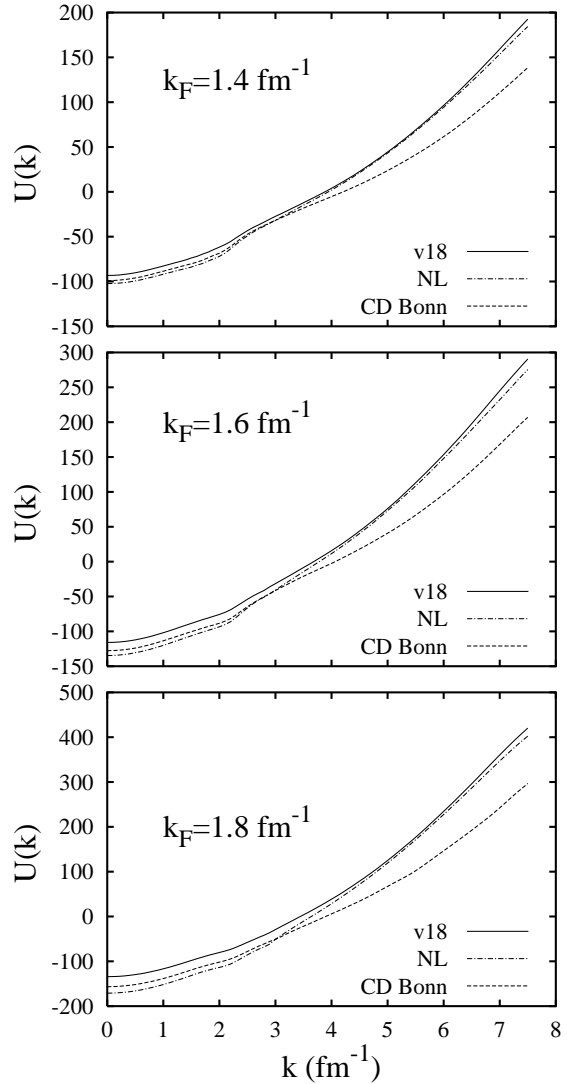}
\caption{Potential $U(k)$ (MeV) of eq.~(\ref{eq:u}) for $k_F=1.4$
(upper panel),
$1.6$ (middle) and $1.8$ $fm^{-1}$ (lower) and for the  v$_{18}$ (solid line),
non-local (dot-dashed) and CD Bonn (dashed) potentials.}
\label{fig:upot}
\end{figure} 
%
In order to get further insight on the origin of the differences
in the potential binding energy, in fig.~\ref{fig:upot} we also 
show the potential $U(k)$ for three different choices of the density,
namely $k_F=1.4$, $1.6$ and $1.8$ fm$^{-1}$.
The most important contribution to the total binding energy
comes from the low $k$ region, where the NL potential is, indeed, lower
than the others. At large $k$ the behavior of the CD Bonn curve differs
from the others, which is not surprising considering the different
high momentum tail of the potential $V$ shown, for example, in
fig.~\ref{fig:diag}. However this behavior does not seem to affect
the final result for the binding energy.  
\par
\begin{figure}[t] 
\includegraphics[width=0.98\columnwidth]{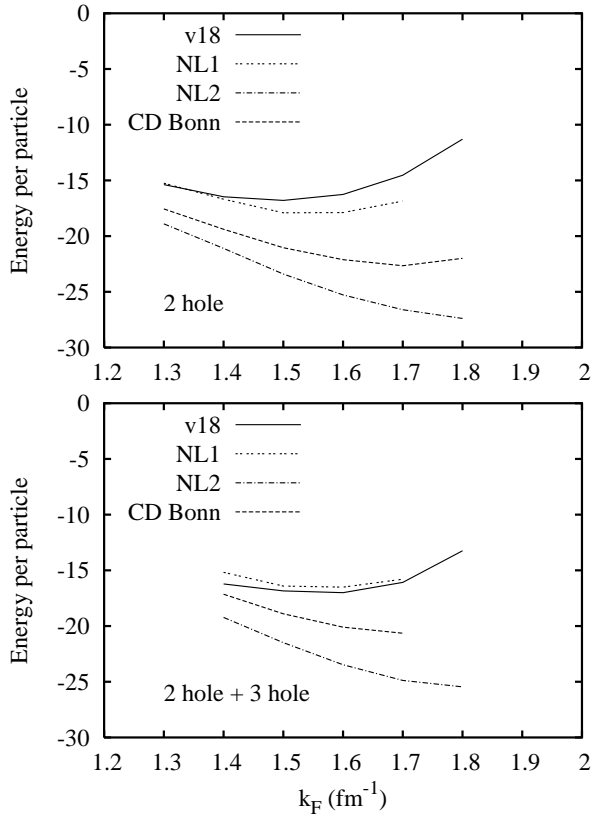}
\caption{
Energy per particle (in MeV) for symmetric nuclear matter calculated 
for the v$_{18}$ and CD Bonn potentials and for the non-local 
interactions NL1, which includes the non-locality
in the $^1S_0$ channel only, and NL2, where the non-locality
is included both in the $^1S_0$ and in the $^3S-D_1$ channels.
The upper panel shows results at the two hole line level, while in the
lower panel three hole-line contributions are included.}
\label{fig:sat}
\end{figure}

Finally, 
the saturation curve of nuclear matter is shown in figure~\ref{fig:sat},
where, in order also to make contact with the existing literature, {\it e.g.}
\cite{Engvik:1997er,mach98,mach96}, in the 
upper panel we show results at the Brueckner level, while in the lower panel
we include contributions at the three-hole line level.

Again, as a reference curve we take the one obtained using 
the Argonne v$_{18}$ interaction (solid line):
as it is well established, the saturation density it gives 
is slightly larger 
than the empirical one, and three-body forces can be introduced to 
correct for this drawback \cite{book,tbf}. 
This is just the issue we want to
address, and therefore all the calculations will be presented without
the introduction of three-body forces, in order to see to what extent
the non-locality can affect this result and in which direction. 
The discrepancy of the calculated saturation point from the empirical one 
can be taken as a measure of the amount of three-body forces which is needed.
Also relativistic effects can shift the saturation point, as
described within the Dirac-Brueckner (DB) approach \cite{obe}.
However, it has been shown in ref. \cite{BrownWeise} that the
relativistic effects introduced in DB calculations can be
described in terms of non-relativistic three-body forces due to
virtual creation of nucleon-antinucleon pairs.  
\par
Let us now consider the other two interactions, which display different
degrees of non-locality. The CD Bonn at the Brueckner level has been already
studied in ref. \cite{bonn_old} and more recently in ref. \cite{bonn_new}.
The corresponding saturation curve reported in fig.~\ref{fig:sat} 
is in agreement 
with previous studies. The saturation density and energy are substantially 
larger than for v$_{18}$. 

The introduction of three hole-line
contributions slightly improves the situation, since the three hole-line 
diagrams give an overall additional repulsion, 
which however is much smaller than
the two hole-line contribution. As shown in the bottom panel of
fig.~\ref{fig:sat}, the binding energy decreases, but still remaining
too large with respect to the empirical value, while
the saturation density is essentially unchanged.
All that indicates that, in order to get 
the correct saturation point, one needs an amount of
three-body force larger than in the v$_{18}$ case. This is somehow at 
variance with the results for few nucleon systems, where the CD Bonn seems to
produce binding energies closer to the experimental ones \cite{cd_few}.
\par

The Doleschall' s potential moves further in this direction. 
As in fig.\ref{fig:bind} we consider again the NL1 and NL2 cases.
%
%
The saturation
density, as one can see in fig.~\ref{fig:sat}, is larger than in the
v$_{18}$ case already for the NL1 curve and it becomes abnormally 
large when both $^1S_0$ and $^3S$-$D_1$ channels
include non-locality (NL2), with the
energy per particle going down to about -26 MeV. The very large strength
for the three-body forces needed in this case, for shifting the
saturation curve close to the phenomenological one, is in sharp
contrast with the excellent results obtained 
\cite{Doleschall:2000rp,Doleschall:2003ha,Doleschall:2004bb,
Lazauskas:2004hq} for few body
systems, where virtually no three-body forces are needed. \par

It looks that any increase of the non-locality would improve the
fitting of the binding for the few body, but would shift the saturation point
to higher density and binding energy. 
\par
\section{Conclusion}
\label{s:concl}
We have analyzed the effects of the non-locality of the NN interaction on
the saturation curve of nuclear matter by considering a set of 
interactions with different degree of non-locality. While the introduction
of non-locality is able to improve the agreement with phenomenology in
few-body systems, where it can reduce or even eliminate the need of three-body
forces, in nuclear matter we observe a parallel shift of the saturation
point in the wrong direction, away from the empirical saturation point.
The presence of non-locality seems to be unable to solve one of the 
fundamental problem in the many-body description of nuclear systems,
i.e. the (in)consistency of the three-body forces needed in few-body nuclear
systems and in nuclear matter. 
\acknowledgments
We thank P. Doleschall and R. Machleidt for providing the subroutines
for the non-local and, respectively, the CD Bonn potentials. We would
also like to thank P. Schuck for 
illuminating discussions on the problem
of non-locality, which stimulated the present work.

%
\end{document}